\documentclass[sigconf]{acmart}

\usepackage{graphicx}
\AtBeginDocument{%
  \providecommand\BibTeX{{%
    \normalfont B\kern-0.5em{\scshape i\kern-0.25em b}\kern-0.8em\TeX}}}

\setcopyright{acmcopyright}
\copyrightyear{2025}
\acmYear{2025}
\acmDOI{10.1145/3706598.3713940}

\acmConference[CHI 2025]{CHI Conference on Human Factors in Computing Systems}{April 26--May 1, 2025}{Yokohama, Japan}

\acmPrice{15.00}
\acmISBN{979-8-4007-1394-1/25/04}

\begin{document}

\title[The Centers and Margins of Modeling Humans in Well-being Technologies]{The Centers and Margins of Modeling Humans in Well-being Technologies: A Decentering Approach}




\author{Jichen Zhu}
\orcid{0000-0001-6740-4550}
\affiliation{%
  \institution{IT University of Copenhagen}
  \streetaddress{Rued Langgaards Vej 7}
  \city{Copenhagen}
  \country{Denmark}
}\email{jichen.zhu@gmail.com}

\author{Pedro Sanches}
\affiliation{%
    \institution{ITI/Larsys}
    \country{Portugal} 
 }
   \affiliation{%
 \institution{Umeå University}
  \city{Umeå}
   \country{Sweden}
   }

\author{Vasiliki Tsaknaki}
\affiliation{%
  \institution{IT University of Copenhagen}
  \streetaddress{Rued Langgaards Vej 7}
  \city{Copenhagen}
  \country{Denmark}
}

\author{Willem van der Maden}
\affiliation{%
  \institution{IT University of Copenhagen}
  \streetaddress{Rued Langgaards Vej 7}
  \city{Copenhagen}
  \country{Denmark}
}

\author{Irene Kaklopoulou}
\affiliation{%
 \institution{Umeå University}
  \city{Umeå}
   \country{Sweden}
}

\renewcommand{\shortauthors}
{Zhu, Sanches, Tsaknaki, van der Maden \& Kaklopoulou}

\begin{abstract}
This paper critically examines the machine learning (ML) modeling of humans in three case studies of well-being technologies. Through a critical technical approach, it examines how these apps were experienced in daily life (technology in use) to surface breakdowns and to identify the assumptions about the “human” body entrenched in the ML models (technology design). To address these issues, this paper applies agential realism to decenter foundational assumptions, such as body regularity and health/illness binaries, and speculates more inclusive design and ML modeling paths that acknowledge irregularity, human-system entanglements, and uncertain transitions. This work is among the first to explore the implications of decentering theories in computational modeling of human bodies and well-being, offering insights for more inclusive technologies and speculations toward posthuman-centered ML modeling.

\end{abstract}






\maketitle

\section{Introduction}

Human-centered Machine Learning (ML) is an emergent trend in research ~\cite{Riedl2019,Wang2019,Shneiderman2020,capel_what_2023,riener2006human,nguyen2022towards,zhu2020Personalization}. These efforts represent a much-needed departure from the traditional engineering-based approaches, which concern the technical development of ML automation capabilities, often with minimal involvement of human stakeholders \cite{xu2019toward}. 
By comparison, the research area of human-centered ML is committed to developing ML systems that are beneficial to society by emphasizing user interactions, ethics, trust, and other important human factors \cite{capel_what_2023,Riedl2019,xu2019toward}. 
To design and evaluate human-centered ML applications, researchers have drawn theories and methods of understanding humans directly from Human-Computer Interaction (HCI), especially the early waves of HCI \cite{bodker2015third} around human factors and user-centered design~\cite{xu2019toward,jiang2023situation,chancellor_who_2019}. The overarching claim is that ``by engaging with a human-centered process in the design and evaluation, the ML that is designed is inherently human-centered''  \cite{capel_what_2023}.  

Despite its promise, there are signs that design approaches based on human-centered ML can implicitly center certain notions of the ``human'' and their relation to the world. Studies show that human-centered ML often presupposes a superficial universalism of ``humans'' while centering established norms and thus inadvertently marginalizes less privileged groups in algorithmic biases \cite{costanza2018design,chancellor_who_2019,miceli2022studying,o2016weapons,mehrabi2021survey}. 
In addition, by primarily focusing on the {\em direct interaction} between users and ML (through lenses such as usability and human factors), human-centered ML inherited early HCI's conceptualization of the ``human.'' 
It leaves out the complex entangled {\em relations} between human actors and nonhuman actors (e.g., ML-based systems, environment), especially at odds with how ML applications increasingly {\em learn} from their encounter with humans in the world, {\em act}  on their own initiative, and {\em adapt}  themselves through continual updates. 

Current critical discourse on human-centered ML mainly focuses on algorithmic biases associated with data practices (e.g., \cite{costanza2018design,crawford2021atlas,o2016weapons,miceli2022studying}) and humanistic design theories (e.g., \cite{nicenboim2020more,nicenboim2023designing}). What is under-explored is the role of ML modeling practice, that is, the {\em computational process} through which data becomes an ML model of the ``human.''  Critical literature on early symbolic AI has shown that algorithms can encode ideology and values through problem formulation \cite{agre2014toward}. In addition to data, ML models require a technical formulation of a problem to be solved, which includes how the problem is formalized, how the data is processed into an ML model, and what the expected inputs and outputs of the model are. So far, relatively little HCI research has investigated the impact of ML modeling on human-centered ML. A notable exception is Chancellor et al.'s analysis of how ML practitioners represent human research subjects in human-centered ML prediction of mental health \cite{chancellor_who_2019}. Their analysis focuses on discourse in technical publications, not the assumptions entrenched in ML modeling itself.

This paper aims to surface what assumptions of the ``human'' are embedded in well-being technologies, an important domain of human-centered ML \cite{ozmen2023six}. 
Well-being technologies have been developed to assess and evaluate sensed data to manage bodily functions \cite{campo_woytuk_touching_2020, benabdallah2023notebook, lockton2020sleep}, physical performance \cite{rapp_self-tracking_2020}, contraception \cite{park_ambivalences_2023}, fertility \cite{homewood_ovum_2019, costa_figueiredo_self-tracking_2017}, mental health \cite{sas2020mental,sanches2019hci}, as well as chronic conditions \cite{vafeiadou_self-tracking_2021, ayobi_quantifying_2017}. 
These approaches often rely on models of human health and well-being to create predictions of various physiological and bodily processes \citep{reime2023walking, loh2022automated},
opening the gateway to adaptive interventions \cite{sas2020mental} and other forms of personalization \cite{pieritz2021personalised}. As ML-based technologies continue to permeate our daily lives, it is important to ask who they are made for and what conceptions of bodies \cite{homewood2021tracing}, health, and well-being they encode and reproduce. 

To ground our analysis, we select as our case studies three ML well-being apps where users track and manage their lives: 1) Clue, a Menstruation and Fertility Tracking App (MFTA) that predicts menstruation cycles; 2) a social fitness app with personalization ML for motivation; and 3) the Oura ring, a wearable device that classifies well-being states. Inspired by critical technical practice \cite{agre2014toward}, we investigate how the ML features in these apps were experienced to surface breakdowns ({\em technology in use}). We then identify how narrowly construed assumptions about human bodies and lived experiences are encoded in the ML modeling as well as in the design of these apps {\em (technology design}). Next, we decenter the identified assumptions of humans in foundational aspects of those models (e.g., cycle regularity, preconceived notions of users separate from the environment, and health/illness binaries) using agential realism \cite{barad2007meeting} as an epistemological stance. 
From there, we propose alternative centers (e.g., cycle irregularity, human-system entanglements, and uncertain transitions) toward posthuman-centered well-being technologies.  
This paper makes the following contributions:
\begin{itemize}
    
\item We are the first to apply decentering theories to ML modeling of human bodies in well-being technologies. 
We extend HCI critical literature of ML by shifting the focus 
to the assumptions and worldviews of the ``human'' entrenched in the technical problem formulations of ML modeling. Inspired by critical technical practice \cite{agre2014toward}, we incorporate HCI first-person methods to surface breakdowns of {\em technology in use} and use the posthuman philosophical stance of agential realism to de-familiarize and open up technical and design alternatives for each case.


\item We reveal which assumptions of humans and bodies are centered and marginalized in current ML modeling practice through three case studies of well-being technologies. Our analysis shows that common ML features (e.g., classification, prediction, and personalization) privilege assumptions of body regularity and health/illness binaries and downplay the interconnections between humans and nonhuman actors (e.g., ML and the environment). These oversimplified notions of humans in the ML models are in tension with users' bodily lived experiences. 
    

\item We propose a new research direction of posthuman-centered ML in well-being technologies. Using decentering approaches of agential realism, we speculate more inclusive design and ML modeling paths that acknowledge irregularity, human-system entanglements, and uncertain transitions. For designers with various levels of AI literacy, we propose conceptual tools and design considerations to explore more inclusive relations between ML models and human bodies. 
    
\end{itemize}


\section{Background}

\subsection{ ``Humans'' in Human-Centered ML} 

Current human-centered ML research mostly uses methods from the early waves of HCI \cite{bodker2015third} around human factors. For example, constructs such as {\em mental models} \cite{bansal2019beyond,chromik2021think,kulesza2012tell,villareale2022want}, {\em usability} and {\em learnability} \cite{amershi2019guidelines,liao2021human,Shneiderman2020,furqan2017learnability}, and {\em human factors} \cite{auernhammer2020human,xu2019toward} have all found renewed interest in recent years. Similarly, traditional design methods (e.g., user-centered design and end-user testing \cite{xu2019toward}) have been regularly adopted, while more recent methods (e.g., speculative design) were underrepresented in recent ML applications~\cite{capel_what_2023}.  

This gravitation towards early HCI and interaction design methods has a significant impact on how human-centered ML approaches notions of ``human'' and ``human-centeredness.'' 
Researchers have already problematized the inclusiveness of ``the human''  in human-centered ML \cite{costanza2018design,chancellor_who_2019}. With some exceptions (e.g., \cite{ehsan2024xai}), existing work inherited the implicit notion of the ``Universal Human''  \cite{braidotti2019posthuman}, which can lead to exclusionary practices and perpetrate social injustice through algorithmic biases \cite{miceli2022studying,o2016weapons,mehrabi2021survey}. As Estrada \cite{estrada2020human} warns, at its worst, the pretense of inclusive universalism in the uncritical ``human-centered'' rhetoric may ``\textit{be used to justify and provide cover for narrowly self-serving, exclusive, or imperialist practices''  by and for the systematically privileged}''  (p.2). Similarly, Miceli et al. \cite{miceli2022studying} pointed out the power asymmetry with data quality, data work, and data documentation, which can lead to discriminatory or exclusionary outcomes. 

In addition to issues of inclusion, researchers have looked into how the ``human''  is conceptualized. In the domain of ML prediction of mental health from social media feeds, Chancellor et al. conducted thematic discourse analysis to understand the ``human''  in 55 human-centered ML publications \cite{chancellor_who_2019}. They found complex, paradoxical representations of the ``human,'' as both the subject and object of ML, and warned of the risk of dehumanization. Similarly, Ehsan et al. looked at the background of users of ML explanations and found that their ML background shapes users' perceptions \cite{ehsan2024xai}. Built upon our reflections as designer-users and developers, this paper extends this small body of work by surfacing assumptions of the ``human'' in the problem formulations of how ML models human bodies. 

The conceptual framing of the relations between the \textit{human} and \textit{context} is another critical aspect. 
Design efforts of human-centered ML primarily concern the {\em siloed interaction} between users and ML, evidenced by the preoccupations with usability \cite{amershi2019guidelines,liao2021human}, mental model alignment \cite{bansal2019beyond,chromik2021think,kulesza2012tell,villareale2022want}, and human factors \cite{xu2019toward}. Similar to early waves of HCI~\cite{bodker2015third,harrison2007three}, this brand of human-centeredness presupposes that humans are the only actors \cite{jiang2023situation, Shneiderman2020}. This assumption, however, is increasingly at odds with recent HCI and design theories that indicate that humans and technology co-exist in increasingly entangled relations \cite{frauenberger2019entanglement,giaccardi2019histories}. Niceboim et al. \cite{doi:10.1080/07370024.2023.2283535} mention that ``\textit{the dominantly human-centered design approach has been useful in surfacing human needs, but has been less useful in understanding how human concerns are entangled with larger ecosystems and nonhuman agencies}'' (p.3). 
In the context of three ML well-being technologies, this paper provides detailed accounts of how this siloed notion of interaction can cause breakdowns when users interact with them in real-world contexts. 

Our work is also aligned with developments in design research within HCI, where designers depart from seeing the world as a system that can be understood and improved towards transdisciplinary work that opens up “emergent possibilities” \cite{irwin2022transition} and an openness to the complexity and uncertainty of the world. We draw particularly on works that problematize and open up possibilities for how bodies get taken into account in design practice \cite{homewood2021tracing, howell2018emotional, hook2018designing}. Such prior research does not separate between cognitive, on one hand, and physical, emotional, and sensorial aspects of bodily lived experiences \cite{staahl2022making, hook2019soma} on the other. Using performative and practice-based approaches, HCI researchers have also started mapping these concerns into how bodies are modeled and datafied through sensor- and AI-based technologies \cite{sanches2019ambiguity, howell2018emotional, reed2024shifting} and examined how physical physiological bodies are entangled with the world.


\subsection{Posthuman Concepts in HCI and Design: A Focus on Decentering} 

Challenging this predominant human-centered perspective, a growing body of HCI work has drawn on posthuman theories. These design researchers are becoming more attentive to the entangled relations among the multiple actors participating in design processes- including human and nonhuman bodies in the form of data, digital and physical materials, as well as animals, plants, and fungi \cite{10.1145/3491102.3501901,nicenboim2023designing}. A notable example is Wakkary's \cite{wakkary2021things} approach of \textit{designing-with}, in which humans are close collaborators with nonhumans, bound together materially, ethically, and existentially. Similarly, Giaccardi and Redström have called for moving past the “\textit{blind spots of human-centred design, and address the expanding universe of algorithms, forms of intelligence, and forms of life entering design practice, casting them as partners in a more-than-human design practice}” \cite[p.41]{9209270}. For in-depth overviews of posthumanism in design, refer to Forlano's extensive literature review \cite{forlano_posthumanism_2017} and Frauenberger's proposal for Entanglement HCI~\cite{frauenberger2019entanglement}.

In this paper, we draw on the posthuman concept of {\em decentering}. Nicenboim et al. \cite{doi:10.1080/07370024.2023.2283535} articulate how decentering can be manifested in the interplay of posthuman theory and practice, with the goal to ``\textit{recognize and prioritize multiple voices, especially those traditionally marginalized or excluded}'' (p.2). 
They suggest that ``\textit{contextualizing decentering within theory might help designers and scholars [...] to be more precise in their commitments and recognize their limitations. That is, for example, to avoid creating similar blind spots as previous traditions, such as unintentionally establishing new, undesirable centers. 
}'' (p.3). 
Relatively few works have turned to posthuman theories for designing ML applications. Notable exceptions include Nicenboim et al. \cite{nicenboim2022explanations}, whose work aimed to challenge human exceptionalism through more-than-human perspectives and to broaden the notion of “the user” when designing ML-powered conversational agents. Benjamin et al. \cite{benjamin2021machine} analyzed ML artifacts through post-phenomenology theories and argued that ML outputs are ``\textit{inherently characterized by uncertainty from data noise and model variance}.'' 
However, the field of HCI still does not fully understand how posthuman theories could translate to design and development of human-centered ML apps. This paper adds to this small but growing body of work by extending the posthuman concept of decentering to a new area of analysis (i.e., the ML modeling of humans) and new design domains (well-being technologies). 

\subsubsection{Decentering through Agential Realism}

For tackling issues of how ML models are made to represent the world, we draw on Agential Realism \cite{barad2007meeting}, a posthumanist onto-epistemological stance developed by Karen Barad. A representationalist view on modeling sees computational models as capable of representing or providing information about phenomena ``out there'' in the world. Instead, an agential realist account of modeling accounts for how phenomena are created through specific practices of data production and processing in their interrelation with other practices of observation and sense-making. Rather than existing a-priori, phenomena are entangled with the measuring apparatus and created through intra-actions, which are cuts separating the phenomena from the rest of the world. Unlike interaction, which starts with the premise of separate entities and examines actions between them, intra-action emphasizes the inherent entanglement of matter, including the relationship between agencies in an act of observation.

There is a growing interest in agential realism in HCI to diffractively rethink measurement, data, and interaction design as intra-actions across agential cuts \cite{barad2007meeting,barad2018diffracting} in ongoing phenomena. Agential realism has been introduced in HCI to explore new design approaches in a world where technologies---like ML, cyber-physical systems, extended reality, and neuro-implants---are increasingly entangled with our bodies, challenging traditional views of agency and clear distinctions between humans and machines \cite{frauenberger2019entanglement}. Agential realism has also been adopted in soma design to highlight the mutual shaping of designers' bodyminds and materials \cite{validity_rigor_2021} and in contexts where design artifacts become deeply intertwined with users' daily lives over time \cite{making_new_worlds}.

More closely related to this paper, agential realism has been proposed to guide the design of data-driven design artifacts and computational systems aimed at modeling aspects of human behavior. An example of this can be found in the work of Lupton and Watson \cite{lupton2021towards}, where they work with agential realism to inspire collaborative design methods to elicit feelings, practices, and imaginaries concerning living with digital data. Howell and colleagues \cite{howell2016biosignals}, focusing on biosensing, ask what if designers stopped viewing data as an abstract, insight-rich ``thing'' and instead focused on the ever-changing nature of materials and meaning, addressing data's materiality in the design process. Sanches et al. \cite{sanches2022diffraction} import the notion of a diffractive practitioner \cite{hill2017more} to design practice, treating data as a material in interaction design and investigating how designers attribute meaning and understanding to human bodies through diffraction-in-action. Their work shows that living with data, over time, allows designers to see it from a simple representation to a dynamic process, where meaning arises from the entanglement between data and the world. Going a step further, Reed and colleagues \cite{reed2024shifting} show how designers of digital music instruments encode specific symbolic and music-theoretical knowledge in their artifacts, which they analyze as Baradian apparatuses to produce phenomena as sound. They show that musical instruments work as apparatuses with specific assumptions encoded in them, e.g., gestures required to play them, as well as forms of interaction that may have been left implicit or explicit. These generate or reconfigure ambiguities when played by musicians, who may or may not share the same conceptual frames as the designers. They call this a ``productive indeterminacy,'' allowing for experimentation. 

In this paper, we extend prior work by applying an agential realist lens to the ML modeling human bodies, behavior, and experiences in well-being technologies. We show how an agential realist lens, when applied in conjunction with a critical technical practice, can be used to decenter assumptions made by designers and find novel ways to model aspects of human well-being. Like in prior work \cite{sanches2022diffraction, reed2024shifting}, we analyze how systems are interpreted by end-users when used in everyday life and specific contexts of use. And like Reed and colleagues \cite{reed2024shifting}, we highlight assumptions and aspects centered in the design of these technologies. We also extend the agential realist lens to the technical practice of ML development to examine how the relations between humans and technology are encoded in ML algorithms.
We go a step further in conceptualizing how agential realism can be used to illuminate paths forward to decenter which notions of the ``human'' are modeled. 



\subsection{Critical Technical Practice}
Our overarching methodology is critical technical practice. Initially developed by Agre in symbolic AI research \cite{agre2014toward}, critical technical practice uses critical tools to defamiliarize ideas in technical practice in order to reveal hidden assumptions reproduced through ``\textit{unconscious mechanisms such as linguistic forms and bodily habits}.''  
Agre's approach has been adopted in a range of reflective technical practices in AI  \cite{mateas2001expressive,zhu2009intentional,harrell2013phantasmal}. 
In HCI, critical technical practice has also been adopted from early on under the term Reflective HCI \cite{CTP_reflective}. It inspired Sengers et al.' proposal for critical design where they invite researchers to reflect on the centers and margins of design practice \cite{sengers1999practices}. More recently, reflective HCI has also inspired an approach by \citet{howell_auditing} for a participatory AI auditing aiming to identify breakdowns in emotion classifiers. These orientations to design and computer science have additionally been the grounds to practice anti-racist HCI \cite{CTP_antiracist}, to analyze how disciplinary and institutional structures influence impact in computer science work \cite{CTP_seamwork}, and to recover ``lost''  ideas from the history of ML \cite{CTP_strangerationist}. It has also been used to propose a decolonial ML \cite{mohamed2020decolonial}, to align technical ML research with social values 
and reflective explainable ML \cite{ehsan2021expanding}. 
Adding to this body of work, we combine critical technical practice with decentering through agential realism---our research rests epistemologically on agential realism, through which we conduct our reflective critical technical practice.


%
 

\section{Methodology}

This paper follows the invitation of \citet{van2024pluralising} that, rather than formalizing a particular set of methods and tools for conducting critical technical practice, Computer Science and HCI researchers can draw on a plurality of methods and approaches \cite{CTP_reflective, querubin2024climate, sengers1999practices,vertesi2016engaging}. 
Below we articulate our collection of methods, tailored to each of the three case studies' specific contexts.

\subsection{Selection of Case Studies}
We selected three projects for our case study analysis. 1) The menstruation and fertility tracking app \textit{Clue} that uses ML to {\em predict} menstruation cycles. 2) A fitness app that uses ML to {\em personalize} a user's social environment for motivation. 3) The \textit{Oura} self-tracking ring that uses ML to classify a user's well-being state. We selected these particular technologies for several reasons. Firstly, they relate to research projects in which at least one of the authors has been involved. 
Secondly, they cover a variety of well-being self-tracking technologies and ML modeling techniques. These selected cases speak to diverse aspects of user modeling (i.e., of human physiological processes, preferences, and behavior) (Table \ref{tab:overview}). We acknowledge that not all aspects of well-being or human populations are represented here. However, these case studies are intended to surface assumptions on human inclusivity in ML-based well-being technologies in relation to individual bodies (case study 1), individuals in relation to others (social comparison factors) (case study 2), and individual bodies in relation to aspects of time and context of use (case study 3).

\begin{table*}[ht!]
\caption{Overview of the three case studies of well-being technologies.}
\label{tab:overview}
\centering
\begin{tabular}{cp{14em}p{14em}p{14em}}
\toprule
\textbf{\#} & \textbf{Well-being Technology} & \textbf{Key ML Features} & \textbf{Subject of ML Modeling} \\
\midrule
1 & Menstruation and fertility tracking & Prediction of periods, ovulation, and PMS & Physiological processes of individual bodies \\
2 & Social fitness for activity tracking & Personalization of social comparison environments & Psychological preference for social comparison \\
3 & Self-tracking wearable & Classification of well-being states & Behavior in relation to time and context of use \\
\bottomrule
\end{tabular}
\end{table*}

\subsection{Analysis and Methods in Each Case Study}
Our strategy for engaging with critical technical practice was to work with methods that allowed us to articulate aspects of \textit{technology in use}, in combination with aspects of \textit{technology design}, to surface breakdowns in modeling. The experiences of \textit{technology in use} were gathered through first-person methods \cite{desjardins2021introduction}, where authors themselves combine the roles of practitioner and user, as part of a reflective and critical design practice. In case study 1 we use the walkthrough method \cite{doi:10.1177/1461444816675438} to surface user experience aspects of a well-being application. This extends prior research in which the third author was involved \cite{reime2023walking}, including an analysis of menstruation and fertility tracking apps to surface bodily normativities. In case study 2, a more classical approach of critical technical practice was taken \cite{agre2014toward}, where the first author identifies discrepancies between technical decisions and user study results. She then relies on critical social science and humanities theories to problematize assumptions around personalization. In case study 3, the fifth author adopts an auto-ethnographic approach \cite{kaltenhauser2024playing} to surface long-term aspects of using the specific technology. Overall, we have opted for a pluralistic approach to identifying breakdowns in the context of technology use and development in order to adapt to the specificities of each case study.

Characteristics of \textit{technology design} were articulated through analysis of technical documentation and publications of the models and the corresponding design elements, i.e., how and what type of human data it aggregates, how it models such data, and how the models are utilized in relation to the user. Based on this information, we critically reflected on the entrenched assumptions and values in the technical and design practices, identifying what notions of ``humans'' and ``interaction'' are {\em centered}. 
Next, we attempted to decenter the identified assumptions using agential realism as an epistemological stance. We speculated on alternative design and ML modeling paths aiming to make these technologies helpful for all humans, while at the same time surfacing the entangled relationships between humans and modeled data in each case. 
The initial analysis of each case was conducted by the respective author(s) involved in each project. After this initial phase of the analysis, all authors discussed the findings of each case study collectively and contributed with more nuanced details on identified assumptions and potential breakdowns these might cause to user experience. Finally, all authors hosted collaborative sessions focused on decentering and identifying alternative ways of conceptualizing modeling in each case study. 

Below, we reflect on possible impacts of our positionality within the research apparatus in problem formulations and methodology. 
All authors do design research and technological (software/~hardware) development in a Western ethnographic context. Thus, we all have access to public parks, gyms, sports facilities, and reproductive health services, and we all live where taboos around menstruation and reproduction are not prohibitive. This also shapes the types of well-being technologies we worked with and selected for this analysis. The selection of ML well-being technologies and our ability to study them from a user perspective are shaped by our location in that we are among the intended users, e.g., as English-speaking persons living in areas with access to reliable technological and/or health and sports infrastructures. Additionally, all authors work full-time in academic settings where ideas of Computer Science and Informatics are dominant, which can often reinforce the hierarchies of knowledge and sense-making that our work aims to surface, unpack, and critique. We chose first-person approaches for our analysis because they enable us to identify nuanced cases of mismatch between modeling and lived experience, which would be difficult to assess otherwise by means of alternative qualitative methods, such as interviews \cite{desjardins2021introduction}. However, this also entails that our analyses are integral to our own reflective and critical design practice and that we are only able to correctly identify breakdowns anchored on the authors’ lived experience. Therefore, our results are not generalizable, nor do we claim completeness. At its core, our work points out that even among intended users, there are important aspects of lived experience that could be modeled otherwise. As our findings are rooted in the specific design contexts of the three cases, it is foreseeable that other cases may produce different findings. However, our aim is to demonstrate how our approach can lead to alternative conceptualizations of ML modeling of human.

\section{Three Case Studies on Designing ML Well-being Technologies}
For each case study we firstly present a brief context and background, followed by the reflexive analysis we conducted to identify implicit assumptions on how aspects of human well-being are modeled by those ML and data-driven systems, and how they become prevalent in technology development and use (i.e., to identify the ``centers'' and the ``margins", meaning what is prioritized and what is secondary to the modeling). Then we use agential realism to speculate on new approaches on modeling humans when designing the respective technologies (i.e. decenetering the centers re-centering the margins). 

\subsection{Case Study 1 - Modeling of Bodies in {\em Clue}}
\label{sec:case1}

\subsubsection{Context/Background}
For this case study, we drew on the third author's prior research that used the walkthrough method (e.g., \cite{doi:10.1177/1461444816675438}, \citep{reime2023walking}) for performing a critical app analysis. For this analysis we focus on data that relate to how the app models bodies, and how a user experiences the app's modeling predictions. According to \cite{doi:10.1177/1461444816675438}, the walkthrough method consists of three stages: entry, everyday use, and exit. The first stage of ``entry'' refers to the detailed observation
and articulation of the initial steps made when entering an app, meaning when registering for it. The aim of the second stage, ``everyday use", is to replicate the intended everyday use of an app as realistically as possible while focusing on recording the functionality, options, and
affordances that the app provides to users. The third stage includes the closing/exiting. It focuses on zooming into and reflecting on how users can exit an app, including if, and how, they can delete their account and what happens to the users’ data once they delete an app. 
In this paper, we focus on findings from the walkthrough we conducted on the \textit{Clue} MFTA, due to its particular focus on the prediction of periods, ovulation, and PMS \cite{ClueMainWS}.

Prior work in HCI (e.g. \cite{fox2019vivewell, ciolfi2023analyzing, moniz2023intimate}) have done compelling critiques of over-emphasis on individual empowerment for health, or an over-emphasis of knowledge-seeking through data as a basis for caring for oneself, assumptions that many MFTAs are based on. Here, we leave these assumptions mostly unchallenged for two reasons. First, previous research suggests that a main reason for users to engage with MFTAs is to prepare themselves for different phases of their cycle, such as bleeding and fertile phases, and thus there is a use for these kinds of systems. Second, the scope of our paper is on modeling. Therefore, we focus mainly on the data and the assumptions the modeling itself is based on. 

\begin{figure}[t!]

\includegraphics[width=0.85\columnwidth]{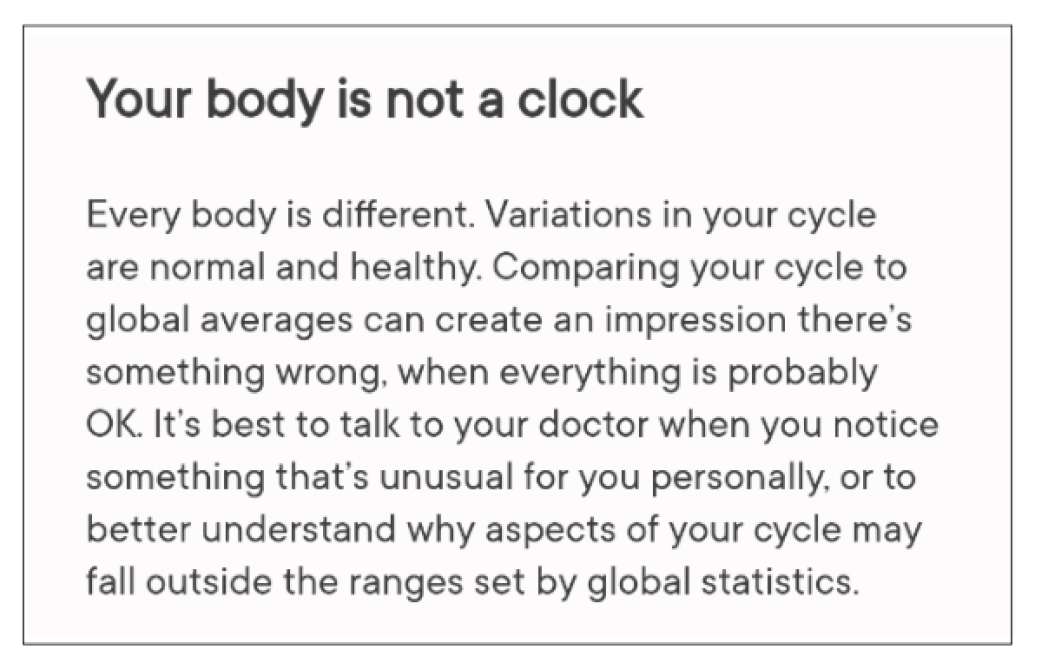}

\centering
\caption{A notification in {\em Clue} stating that comparisons \\ to ``normal'' are based on global statistics (Case Study 1).
}
\label{fig:caseStudies_1}
\end{figure}

\subsubsection{Analysis to Reveal Embedded Assumptions}

Clue promises its users to ``\textit{get predictions based on the most up-to-date science}'' \cite{ClueMainWS}, thus building on the assumption that a crucial factor in intimate health for people who menstruate is the ability to predict and manage menstrual cycles. As the analysis of conducting the walkthrough method has shown, it is not explicit to the user which data is used in their models to predict their reproductive journeys. 

When turning the predictions of the Clue app on, in the information field popping up, it is stated: \textit{“If your cycle length varies, you might want to turn off Clue’s predictions. There might be a higher chance that your cycle length varies, when you are a teenager (younger than 20); you are or recently have been pregnant; you’re breastfeeding, you are in transition age (45 year or older); you are suffering from something that can affect your menstrual cycle (e.g., PCOS or endometrioses).”} Here, it is visible how one assumption embedded in the design of Clue is the \textit{centering of regularity in cycles}. This makes one think that Clue's algorithm can only account for a
certain kind of body, namely a body that averages global statistics (Fig.~\ref{fig:caseStudies_1}). However, Clue is not clear about which specific algorithm is used in predictions, but explicates that \textit{''predictions are based on data from your last 12 cycles, and averages on data from your last 6 cycles''} \cite{CluePredictionWebsite}. 

Additionally, Clue recommends hiding irregular cycles from the data and tracking consistently to improve results. This casts users as regular data providers in order to improve a model that requires regularity for accurate predictions. Prior research involving a dataset of 1.6 million menstrual cycles from 267,209 individuals in the US, all over 18 years old, found that 22\% of individuals experience cycle irregularity \cite{doi:10.1073/pnas.2113762118}. Notably, 99\% of the sample identified as women, which may skew the data towards cisgender individuals and \textit{excluded} teenagers, who are more likely to have irregular cycles. Certain demographics, such as those with chronic illnesses or those at transitional ages, were more prone to irregular cycles, aligning with findings from the Clue app. However, there were also significant ethnic and cultural variations, potentially influenced by a range of cultural, behavioral, and environmental factors that remain largely understudied. As such, irregularity is not, like the modeling may suggest, a rare outlier but rather a central feature of the lived experience of fertility for a large population. We wish to be fair in our analysis and also state that the prediction of regular cycles is a very common feature across MFTAs; therefore, Clue is not an outlier. Additionally, Clue is working with researchers in studies that intend to cast more light into the sources of cycle irregularity \cite{ClueResearchWebsite}, but our point still stands.

We have also identified a particular assumption in the modeling of cycles, which is \textit{centered on a cisgender understanding of fertility}. 
The issue of essentialism in gender-input forms is well-recognized in HCI research \cite{10.1145/3461778.3462033}. This concern is further highlighted by the lack of specification for sex and gender in the dataset used to train and test models \cite{buolamwini2018gender}, with no efforts made to augment the training data to address this gap. In addition to the focus on regular cycles, this centering of cisgender experiences as the normative assumption behind the modeling means that the model will likely not work for individuals undergoing any form of hormonal replacement therapy, in particular hormonal replacement therapy known to cause individual forms of irregularity to fertility cycles~\cite{10.1145/3678575}.

\begin{figure*}[t!]

\includegraphics[width=0.7\linewidth]{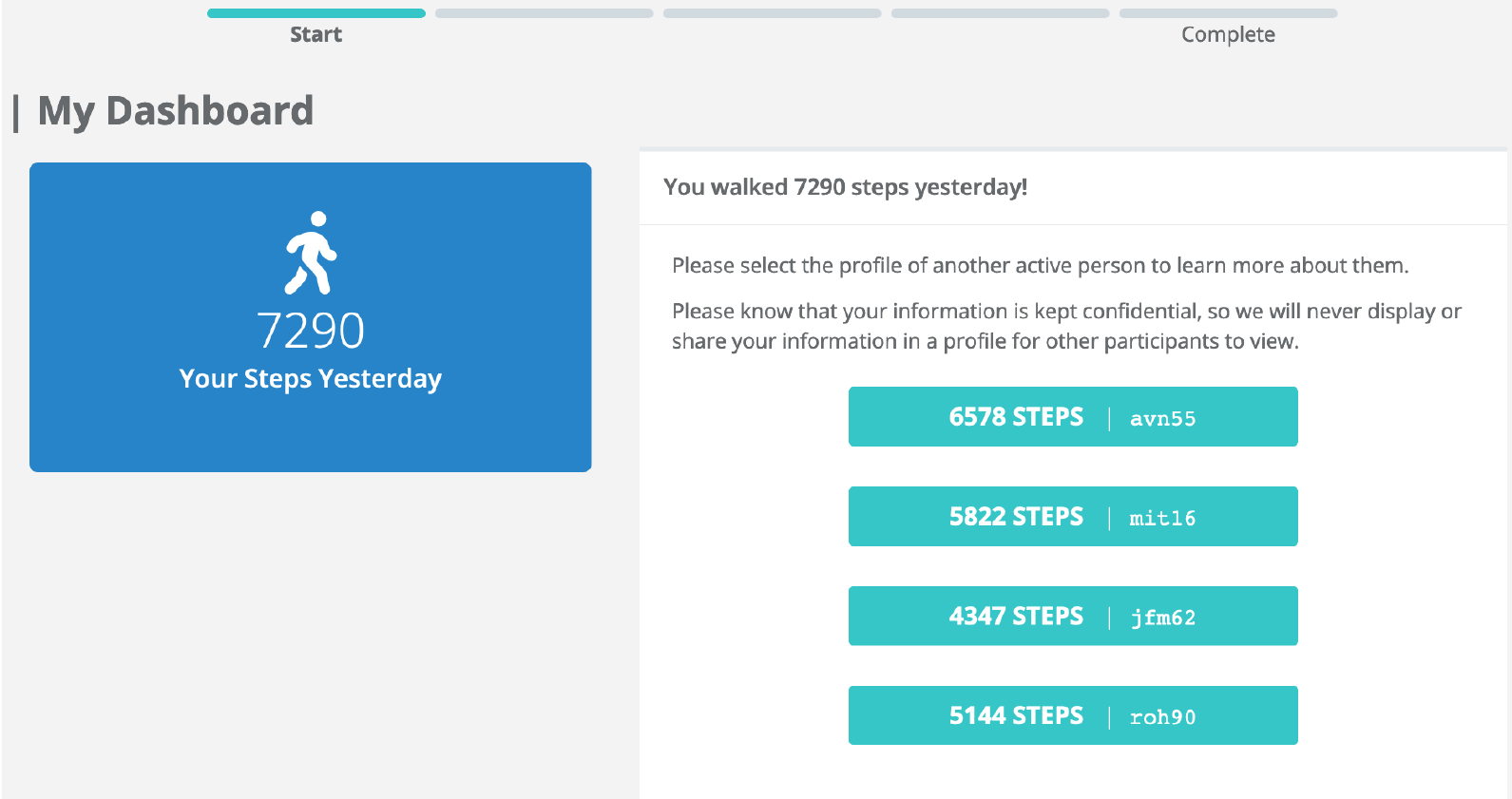}

\centering
\caption{
 Screenshot of a personalized social comparison page with other users to motivate physical activities (Case Study 2). 
}
\label{fig:caseStudies_2}
\end{figure*}

\subsubsection{Decentering Assumptions Through Agential Realism}
Here we speculate on alternative design paths through decentering the assumptions identified in the analysis: individual cycle regularity and cisgender understandings of fertility. These assumptions are centered on the idea of a Human who is a cisgender woman, and who experiences regular and predictive cycles. To decenter these ideas of the users, we turn to agential realism. 

As articulated in the Background, agential realism challenges representationalism. In our case, it allows us to approach modeling not merely as a representation of an external phenomenon --- the idea of a regular cycle. Instead, the world and apparatus--data input forms, the algorithms that predict regular cycles, and the ways that the results of these predictions are communicated to users--create the phenomenon and experience of a ``normal'' and regular fertility cycle, which may or may not be the experience of the users of this app. Instead of letting users compare their experience to a norm, one could design applications that facilitate users' self-exploration through discovering contexts that contribute to different temporalities in menstrual rhythms. For example, by building on research in menstrual irregularity, one could allow users to explore which factors such as hormonal intake, social factors such as rotating shift work, pesticide exposure, racism, and stress can have biological effects on menstrual patterns, as well as different rates of chronic illnesses, such as polycystic ovary syndrome \cite[pp. 11-12]{doi:10.1073/pnas.2113762118}. But also by emotional and lifestyle factors like sex drive, happiness, energy levels and other.

Approaches to modeling bodies in MFTAs can take on many different forms. For example, data collection performed by users, which is used to train system predictions, could facilitate self-discovery of what lies at the heart of differences in cycles. This can be a feature used for modeling bodies and for highlighting inclusive and plural approaches in the ML modeling design. ML-based approaches tied to data visualization techniques, such as cluster exploration, could also be deployed to help users explore how their data may be classified in different clusters and what makes each cycle different. We also see some of these ideas explored in the design of \emph{Drip} \cite{Drip} and \textit{Self} \cite{selfProject}, allowing for experimentation and self-discovery based on self-defined data. By centering users at the margins and by not aiming at achieving a global generalizable prediction of regular cycles, novel ML-based modeling approaches can instead ask the questions: \textit{Can we create cyclical metrics that are appropriate for different sexual and gendered expressions and lifestyles? What social/cultural factors should be included in the pluralistic modeling of cycles? Can irregularity be a feature?}

\subsection{Case Study 2 - Modeling of Preferences in a Personalized Social Fitness App}
\label{sec:case2}
\subsubsection{Context/Background.} 

This case study analyzes the first author's recent project, where a highly interdisciplinary team of ML, HCI, and health psychology researchers investigated how to develop a personalized social fitness app to motivate physical activities\cite{zhu2021personalization,gray2020player}. Social comparison is the psychological process by which individuals evaluate themselves relative to others~\cite{festinger1954theory,wood1989}. 
Existing fitness apps often use social features to tap into this process for motivation ~\cite{olander2013most, leahey2010effect}. Widely used social features such as ranked leaderboards and competitive challenges expose users to more active members in the app for this purpose. While some individuals benefit from this type of {\em upward comparisons} to those who seem ``better off'' (e.g., more active) than them, researchers have shown that others can be negatively affected by it. For the latter group, the opportunity for {\em downward comparison} to those who are ``worse off'' is more motivating \cite{arigo2020social}. Therefore, the project aimed to develop an ML fitness app that provided personalized social environments that suit each user's social comparison preferences in a way all users can benefit \cite{gray2020player,gray2021multiplayer}.  

In this project, ML modeling is used for {\em personalized adaptation}. It is built upon an established approach to automatically adjusting a digital application in order to meet individual users' specifications, needs, or preferences \cite{zhu2020Personalization}. Research has shown personalized adaptation can have a positive effect on an application's learnability~\cite{furqan2017learnability}, usability~\cite{hook1998evaluating}, and user enjoyment~\cite{yannakakis2013player}. 

This project had a substantial technical component of developing new ML modeling approaches to deliver the user experience and health intervention designed by the HCI and health psychology team. In contrast to the rest of the case studies in this paper, its focus is not on breakdowns in users' experience with ML. Instead, it offers us the opportunity to deep dive into the {\em technical practice of ML modeling} and use the method of critical technical practice to analyze how technical assumptions in ML personalization shape users' relation to the technology. We then speculate what personalization may look like if we decenter some of these assumptions.

\subsubsection{Analysis to Reveal Embedded Assumptions} 

The technical practice of ML personalized adaptation typically consists of two main steps: user {\em modeling} to classify individual users and automatic {\em adaptation} of certain aspects of the digital environment based on the classification of a given user \cite{zhu2020Personalization,hirsh2000learning}. 
To model users' social comparison preferences, the ML model tracks the following data from a given user on a daily basis: 1) what type of user profiles they choose to compare with (Fig. \ref{fig:caseStudies_2}) --- do they choose to compare upward or downward, 2) how many steps they make according to their {\em Fitbit} tracker, 3) how they rate their motivation for physical activity. This data is used to train an underlying Multi-Armed Bandits (MAB) ML model~\cite{Auer2002} that learns the user's social comparison preference\cite{gray2021multiplayer,gray2023improving}. 
For instance, if a user is regularly more active and feels motivated when they are exposed to people with more daily steps, the ML model may learn that the user has an upward comparison orientation. And vice versa. Once the ML model reaches a certain level of confidence, the fitness app can then {\em adapt} the social environment and create more opportunities for upward comparison by exposing the user to other users who are more active, as the case in Fig. \ref{fig:caseStudies_2}.

The project in Case Study 2 follows the established ML personalized adaptation technical process in which the AI first models users' preferences in one digital environment and then adapts one's environment based on how the user model classifies that particular individual (e.g., upward or downward comparison preference). When we (the authors) critically reflected on how the above-mentioned ML modeling process functions in relation to the users' fitness behavior, we noticed several problematic assumptions. 

First, the technical decoupling of modeling from adaptation as separate steps in personalized adaptation centers the conceptualization of {\em humans in the void (as the sole actors)}. It assumes that humans are independent of their environment and can exert agency on the environment without being influenced by the latter. Under this assumption, it is sensible to model a person's preference in one environment and then change the environment without questioning whether their preference remains the same. For instance, the user is initially exposed to all types of social comparison environments (e.g., all upward comparison, all downward comparison, or mixed) with similar probability. This way, the ML model can compare the user's behavior in these environments and learn that, for example, the user is more motivated to exercise when exposed to downward comparison. When the app adapts to this user's downward comparison preference, it becomes apparent that the user is no longer in an environment where all types of comparisons are equally likely. Instead, the user will likely be in a new environment where others are frequently less active than them. It is unclear, in this new environment, whether this user will have the same social comparison preference. 

The conceptualization of {\em humans in the void} entrenched in the ML technical practice, however, indicates it is okay not to ask such questions. It assumes that humans exist devoid of their environment and that the environment does not have agency over human preferences. 
In practice, it turns ML personalized adaptation into a paradoxical problem. As humans and their environment are entangled with each other\cite{frauenberger2019entanglement}, making a new environment to match a person's preference in the previous setting is like following a moving target, leading to breakdowns in users' experience when the mismatch between the user and the user model of the personalization AI is too significant. 

Second, the {\em development process} of ML technologies for health behavior-change is at risk of {\em centering humans as a passive recipient of technology}. 
In reality, health behavior change is a multi-year-long process with complex factors deeply ingrained in people's lives and social relations \cite{klasnja2011evaluate,prochaska1997transtheoretical}. However, the development cycle of human-centered ML is often shorter. Many technical decisions need to be made early on without data to validate them. Researchers and developers often need to ``guesstimate'' the consequences of these decisions. As the development process continues, it becomes increasingly difficult to question or reverse them.  
For example, at the beginning of the project, the ML development team needed to know which MAB search technique to use for the fitness app. However, developing all of them and comparing the results through user studies is not feasible. As a compromise, the team built ``simulated users'' based on an existing dataset of {\em Fitbit} users' daily steps to mimic the statistical trends of how a person's daily steps may vary over time. We then tested the different MAB search techniques with these simulated users and selected the technique with the best results. 

As illustrated above, these ``technical'' decisions were rooted in a conceptualization, common in many ML technical communities, of humans as passive recipients of (ML) technology. 
It allowed the project team to see users as data to tune the ML model without unfolding complex questions of how users may {\em react differently} to different ML configurations.


\subsubsection{Decentering Assumptions Through Agential Realism}
In this section, we speculate on alternative ML technical practices that can count for the {\em intra-actions} between humans, ML, and the digital environment more holistically. With the lens of agential realism,  the central phenomenon in this case study is ``social comparison.'' The social fitness app discussed here enacts an agential cut between users' social comparison preference and their digital environment. As our analysis above shows, this cut is problematic because it attempts to separate the entangled relations between users and their social environment. 
To fully acknowledge the agency of the environment on the user, we speculate a new ML approach that formalizes the environment as an ``agent'' in similar ways as the user. For instance, we can use multi-agent learning (e.g., \cite{stone2000multiagent}) to formalize the user and their environment as separate ML agents that influence one another in the ways discussed above. 

Another approach to decentering is to enact less established, alternative agential cuts. The fitness app, like most AI personalization apps, is designed with an ML user model that will eventually become a perfect representation of the human user in mind. When this assumption falls short, it can lead to breakdowns in users' experience. What if we apply a new agential cut between the human and the corresponding user model of them? Using the concept of diffraction \cite{sanches2022diffraction}, we can reframe ML user models as diffraction, instead of representation of the human. 
In the context of this case study, the developers could start with the acknowledgment that a user's social comparison preference may change and the ML user model may be inaccurate from time to time. The development team could explore new diffraction-based human-ML relations by selecting new ML techniques that can capture the {\em co-evolution} between the human and their user model. For instance: 
\textit{Can applications be designed to highlight, instead of hiding, the distance between personalization models and individual users? Can this diffraction create a space for reflection, empowerment, and playfulness?} 


\subsection{Case Study 3 - Modeling of Bodily Rhythms and Well-being Transitions in the {\em Oura} Ring}

\subsubsection{Context/Background}
This case study builds on a six-month autoethnographic research of the fifth author examining the lived, introspective experience of using the Oura ring \cite{OuraWebsite}. The autoethnography sought to explore how algorithmic agencies frame the lived experience of well-being. The motivation for wearing the Oura ring was based on the author's energy fluctuations due to the intersection of social aspects of aging, managing health conditions, and mobility frequency due to work.

According to the manufacturer, the Oura ring is designed to translate the ``body's most meaningful messages — sleep, activity, stress, and heart health — to transform how you feel every day.'' Designed and developed in Finland, Oura is rooted in holistic wellness principles. With access to 20 biometrics (15 sensors) and daily assessments regarding bodily readiness, stress state, sleep patterns, and resilience, the wellness tracker detects and augments transitions concerning sleep (e.g., winding down - sleep – awake), bodily states and activity (e.g., active – non-active – recovering), stress (e.g., engaged – stressed – restored), and menstrual cycle (e.g., period and fertility tracking based on body temperature). 

Oura calculates a ``readiness score,'' which models individual physical and mental capacity throughout the day. Rather than being exclusive to this particular tracker, variations of readiness scores are common features in popular trackers such as Garmin \cite{GarminWebsite}, Fitbit \cite{FitbitWebsite}, and others \cite{WhoopWebsite}. The exact details of how each device calculates readiness scores vary but are generally derived from long-term cardiovascular and sleep measurements \cite{Altini2024}. Oura's readiness score considers overnight and long-term activity metrics, sleep, and physiological data (e.g. resting heart rate, average heart rate variability, recovery index, body temperature). It is categorized into four tiers: Optimal, Good, Fair, and Pay attention. The tiers are designed to support users in their daily planning. The Optimal range (e.g., above 85) indicates that users are ready to take on more challenges, whilst scores in the Fair and Pay attention category (e.g., below 70) suggest rest and recovery prioritization. Within the Good tier, users rely on the assessments and metrics of their readiness feature contributors \cite{OuraReadinessContributors}. The daily score calculations study the deviation of chosen metrics over a 14-day window in comparison to their 2-month trend — with the most current 3-5 days having more weight than past data points. According to the manufacturer, the combination of overnight and long-term metrics in calculating the readiness score accounts for short-term well-being transitions such as illness, menstrual cycle, diet adjustment, and stress recovery. However, as we show below, these shorter-term calculations do not account for more complex transitions.

\subsubsection{Analysis to Reveal Embedded Assumptions}
The readiness score is a gradient ranging from 0 to 100 and the feature is designed to support users in distinguishing days based on their physical and mental capacity. However, the app has two binary modes of use: Normal and Rest. The assumption embedded in the application is that \textit{there is a binary state of being well and being sick, each of them resulting in distinct user journeys}. In Rest mode, the app does not recommend daily physical activities but instead provides advice on how to recover. Whilst this may work for most cases, some episodes in the author's autoethnographic journey illustrate discrepancies between lived experiences of well-being and the scores and modes provided by the app. Through the author's autoethnographic notes, the following two entries present cases of unaccounted transitional bodily rhythms in the readiness feature system: 1) re-establishing long-term baseline due to illness and 2) short-term baseline interruptions due to traveling.

\begin{figure}
    \centering
    \includegraphics[width=1\linewidth]{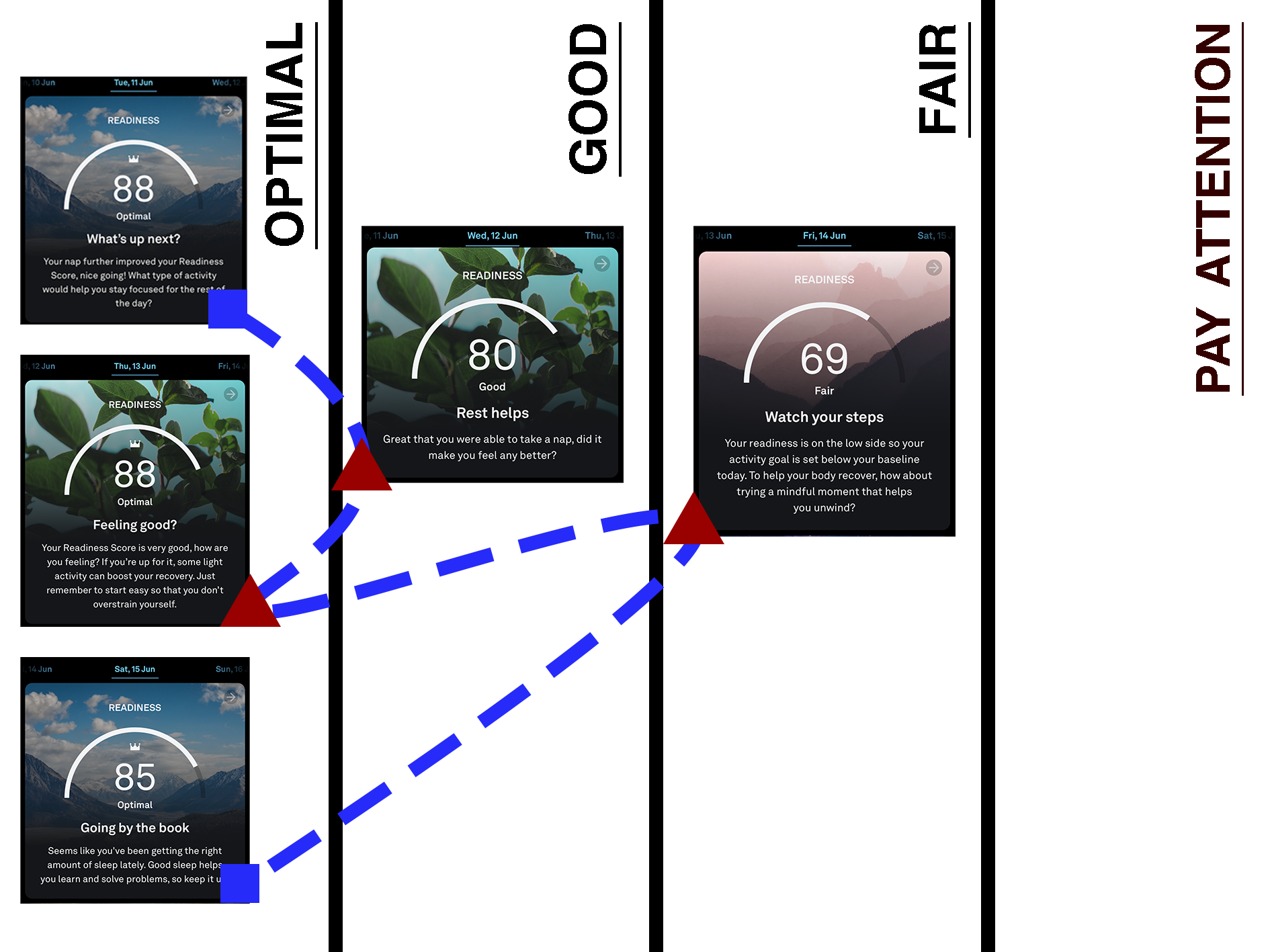}
    \caption{Well-being transitions between readiness tiers and the readiness score assessments in the first autoethnographic entry (Case Study 3).}
    \label{fig:oura-tiers}
\end{figure}

Between March and August 2024, the author experienced periods of unwellness at regular intervals. Those periods included flare-ups of symptoms of dizziness, tremors, vertigo, nausea, chest pain, difficulty breathing, and mild fever. The autoethnographic diary, used for data gathering, indicated that inconclusive healthcare advice was received twice during that period (health services hotline and doctor visit), while the ring illustrated a normal health state. On June 12th, the illness-detection algorithm identified a change in body temperature, despite the high Readiness score, and suggested turning on Rest Mode. They followed Oura’s suggestion for 2 days. The readiness feature demonstrated an overall improvement that influenced them to return to the Normal mode despite not aligning with their subjective, felt health state. The readiness score rebounded to the previous strong metrics after one day as presented in Figure \ref{fig:oura-tiers}. Whereas symptoms improved in the periods in-between flare-ups, their impact on their general well-being established a new, felt sense of normalcy in their ability to respond to daily challenges. This \textit{ephemeral transitional state between being well and being sick} was not accounted for in Oura’s readiness trends. The brevity of unwell periods (5 to 7 days) and the hard-to-predict – often invisible – manifestations of symptoms were either not captured or were diluted in the longer temporal windows of calculation. The author found themselves wanting to manually adjust the readiness score or somehow impact how the algorithm viewed them. This re-occurring episode shows how crucial scores, such as Readiness, in the Oura ring \textit{center on stable bodily rhythms and routines, not accounting for inconsistencies, ephemeral and contextually induced transitions}.

Another autoethnographic entry reveals how the same assumption can manifest in a different kind of transition: traveling. During two instances of traveling for work and leisure, the author became aware of how shifts in location and context also result in lifestyle changes. Moving between different cities, roles, and environments 
created challenges in comparing health metrics and setting appropriate goals. In one city, they were caring for a sick person, while in another place, they were managing a research study. The changes in routine, activity levels, and living conditions impacted the variations of physiological data and activity goals. More importantly, there was always a transitional period of re-establishing balance, routines, and more broadly finding restoration after traveling (without necessarily dealing with rest or jet lag, which are accounted for in the application). The entangled shifts in rhythms, roles, context, and traveling routines entail a new “period” of re-adaptation not currently accounted for in the readiness modeling of wellness and fitness trackers. The annotated travel days in their data exhibit the lack of in-between-stage consideration beyond the binary form of ``good''  versus ``bad''  days strategy. Indeed, the readiness 4-tier assessment system and the binary mode of tracking do not consider the nuances in one's well-being that occur in transitional periods.


\subsubsection{Decentering Assumptions Through Agential Realism}
Our analysis identified moments of breakdown when shorter-term idiosyncratic temporalities and contexts were not taken into account in readiness score calculation. Decentering through agential realism, in this case study, challenges prior ways of cutting-together/apart the agentic qualities of well-being in health tracking and suggests alternative design avenues.

Instead of centering the calculation of long-term, pre-defined individual biographies, a system can be modeled with a focus on moments of transition. Such intra-actions hereby 
demand models that enable the negotiation of multiple and parallel dimensions of temporality in self-tracking.
One could design self-tracking systems that allow individuals to center their experiences around micro and macro transitions, with well-being tracking modes and goals that take stock of their individual bodily rhythms. Models designed for experimenting with various temporal windows---beyond those predefined by the developers---support users to claim agency over the sense-making practices of the systems. 

Resisting the assumption of uniformity in well-being temporalities replaces interaction with a co-evolving scheme of navigating entanglement along with self-tracked sensed metrics. Decentering, in this case, suggests the removal of \textit{auto} from the regulation of bodily states and the rejection of objective normalcy in the modeling of individual physiological and mental capacities. 
Centering on the transitions of bodily rhythms, rather than predefined, rigid stages/states, 
opens the design space of intra-personalized meaning-giving practices.

This case study expands the findings of case study 2 by pointing instead to the modeling assumptions in the temporalities of well-being that affect wellness self-tracking. We would like to go beyond the idea of establishing long-term individual metrics that can change between stages of illness or normalcy and challenge the notion of well-being normalcy in machine learning systems. These can be genuinely helpful, but we would like to complement them with ways for individuals to account for the myriad transitional stages that matter to them. We therefore raise the questions of: \textit{How to address individual bodily rhythms in health well-being when models have pre-defined temporal models? How could we design interaction to increase users' agency over sensed data and algorithmic assessments? How could we support the individual becomings with self-trackers and well-being models? Is it possible to achieve co-evolving models in interaction and ML design?}

\section{Discussion}
\label{sec:discussion}

The three case studies represent well-being technologies with ML features such as classification, prediction, and personalization. In each case, we indicate what is centered and de-centered in the ML modeling. In case study~1, the Clue app centers users with regular cycles and a cisgender understanding of fertility. The ML personalization app in case study 2 centers the contradictory worldview that humans are separable from their environments and passive recipients of personalization technology. Similarly, in case study 3, what is centered is the duality of human well-being, which states between ``well''  and ``sick''  and its uniformity. Thus, at the center of all these apps are over-simplified notions of ``humans'' deprived of their contexts, idiosyncrasies, and complexities. While each case is rooted in the specifics of its design context, they show consistent patterns of how assumptions rooted in the ML models of humans bring tension, when the users interact with these applications to manage their well-being.  

First, ML modeling can encode worldviews through problem formulation. Case study 2 shows how the worldview that humans are separable from their environments is encoded in the user-modeling-adaptation architecture and the Multi-Armed Bandit algorithm. This problem formulation is more than technical details. The designer/developer of this model does the important work of transforming a messy world into a tractable ML technical problem and organizing how we conceptualize ``personalization''  \cite{agre2014toward}. Thus, the users susceptible to changes in their social environment are at particular risk of being misclassified by the ML. 

Second, the metrics used to collect data for these models can further distort the model. All three case studies showed the limitation of the metrics currently used in well-being ML, such as oversimplified ways to quantify the human body, behavior, and preferences irrespective of the context/environment/interactions with the system. As we showed, this limitation of the metrics can lead to various breakdowns with real-world users. 

Lastly, the agency of ML models exerts itself through its entanglement with users. In our cases, discrepancies between users' bodily experience and ML decisions cause users to doubt or re-interpret their experience. Case study 1 has shown how predictive algorithms can often cause users to question or override their bodily sensations when they fall outside the ML prediction. Case study 2 reveals how users' social comparison behavior changes as ML personalization matches them with different individuals. It can lead to a gradual divergence between the user and their model, causing personalization to break down completely. In case study 3, the author experienced friction as she transitioned between the model's pre-determined states. These user reactions to ML breakdowns are not just data points but opportunities for the audience to understand and address, thereby influencing the ML model and preventing further breakdowns. 
Acknowledging the agency, situatedness, and entanglement of ML models is important when designing more inclusive models for well-being technologies.

Our speculations based on agential realism indicate that a posthuman lens can be used to open up new technical ML modeling paradigms. The various agential cuts we proposed above have led to the conceptual foundations for alternative conceptions of the human body and designs that support plurality.

\subsection{Posthuman-Centered ML Modeling}
\label{sect:design_implications}

Our case studies highlight alternative design directions including less rigid, more ambiguous, and self-exploratory predictions. Such design directions aim for different bodies and experiences to be included and feel represented in well-being ML technologies. Ambiguity is a widespread concept in HCI. Researchers have studied it in relation to self-interpretative systems, allowing users to create their own meaning and understanding of data concerning their surrounding or their own body (e.g., \cite{gaver2003ambiguity, sanches2019ambiguity}). In the context of ML, ambiguity and uncertainty have also been investigated as fundamental properties of machine learning as a design material  \cite{benjamin2021machine, sivertsen2024machine, reed2024shifting}. Similarly, in well-being ML apps, the ambiguity of a system's output would necessarily prompt a tracked individual to come up with their meaning-making. Addressing ambiguity would require a shift from prototyping a device that mirrors the body of the user unproblematically to paying attention to relational and experiential aspects of living with data, the different power dynamics involved in data output and visualization (who sees whom), and how user and devices shape each other \cite{tsaknaki2022fabulating, howell2018emotional}. While ambiguity can play an important role in attending to the entangled and relational aspects of modeling and mapping bodily data to different predictions, it is not the full solution. As the case studies show, even systems designed with a level of ambiguity can cause breakdowns and result in inequalities of inclusion/exclusion. We contribute by moving beyond ambiguity, showing instead how a decentering through agential realism can lead to alternative ways of modeling: through identifying ``the centers''  (i.e., assumptions embedded in the process of designing models for well-being technologies) and then decentering those assumptions. 

One aspect that deserves more attention is the co-evolving nature of models in how they represent the world. This aspect is known in machine learning literature as \textit{concept drift }\cite{lu2018learning}, requiring an active approach to updating data and models. In interaction design, research has shown that prototyping with data and ML can fail if it does not account for the evolving and situated nature of data-driven artifacts, where interactions can unfold in unpredictable ways \citep{yang2020re}. Posthuman-centered ML modeling thus requires conceptualizing the entangled relationships of humans and technologies both during the design process and then continuing through the use of the system. This disturbs the neat categories of humans on one side and ML on the other. 
Modern ML-based apps can be said to be constantly in the state of being designed, always evolving through new data \cite{giaccardi2020technology}. Considering complex human-technology relations can help us consider how technology is perceived while simultaneously interrogating how technology shapes humans. 

Another important aspect highlighted by our case studies is the active role of nonhuman agencies in defining what gets modeled, and how modeling is done in practice. Our work exemplifies how engaging posthuman theories can lead to new design avenues that are aligned with the active role of nonhuman agencies entangled with humans. Design researchers who engage with posthumanism have stressed the importance of accounting for the capacities of nonhumans (materials, tools, and software) in a design process \cite{oogjes2022weaving}, as well as the agency of designed artifacts more broadly \cite{giaccardi2019histories}. According to Bennett \cite{bennett2016vibrant}, materials are vibrant through their efficacy as animate entangled things rather than passive individual objects. In that way, all human and non-human matter always depends on the collaboration, cooperation, or interactive interference of many bodies and forces (ibid). 
While human-centered design primarily treats ML and data as neutral, passive technological objects, critical studies of ML and data have pointed out that all technologies encode social values and politics \cite{agre2014toward,crawford2021atlas,zhu2009intentional,o2016weapons}.
ML's increasing autonomy places the challenges of how to design ethical systems through constructs such as machine agency, autonomy, accountability, and creativity at the forefront of HCI design research \cite{Shneiderman2020,heer2019agency,dove2017ux}. Similarly, a posthuman-centered ML modeling approach would be concerned with accounting for the active role of nonhuman agencies (e.g., the role of ML, data and algorithms) entangled with humans. A designer engaging with crafting new ML models in a well-being app should carefully consider how the agency of each socio-material element (at the levels of both discourse and materiality) creates specific conditions for design and hinders others. 

\subsection{Implications}
By surfacing the centers and margins of ML modelings in the three case studies of well-being technologies, we identified some inherent limitations of human-centered ML. In this section, we propose an alternative approach --- {\em Posthuman-Centered ML Modeling} by adopting posthuman theories to reconceptualize ML in the design of ML applications. Below, we describe three conceptual shifts that ML practitioners and designers can adopt to center alternative conceptions of the human body and its relations to technology. 

\subsubsection{From Adding More Data Points to Decentering Epistemologies}
Existing HCI literature has pointed out that ML models regularly misrepresent, if not exclude, marginalized non-confirmative identities and impose a hegemonic epistemology of normative bodies and experiences \cite{costanza2018design}. However, relatively few remedies have been proposed. A common approach is to add data points when modeling human parameters, specifically from those whose bodies and lived experiences are currently under-represented \cite{rajkomar2018ensuring}. Others have also explored increasing diversity through participation (e.g., using participatory design methods, improving diverse representations in the design and development teams) \cite{dennler2023bound, danielsson2023queer, zaga2022diversity}.

We propose that researchers of ML well-being technologies should {\bf surface implicit epistemological assumptions in the ML models they use and decenter those to increase plurality}. While systemic issues like discrimination and diversity fatigue underscore the importance of equitable data collection \cite{smith2021diversity}, we argue that inclusivity must also be embedded within the foundational assumptions of ML modeling. Our case studies show that while equitable data collection \cite{smith2021diversity} can mitigate some issues, it alone may not be sufficient. In case study 1, for example, adding more data points from users with irregular menstruation cycles to the training data of its ML model may shift what the model calculates as ``global averages'' (as in Fig. \ref{fig:caseStudies_1}), but it will do little to decenter the idea of regular cycles. Similarly, in case study 3, more data points from users in transitional contexts alone will not automatically challenge the binary states of ``readiness.'' In order to fully decenter these biases and normative assumptions, researchers need to also operate at the level of ontologies and categories in ML modeling of certain well-being-related phenomena.

Our case studies have demonstrated our methodological approach of combining critical technical practice and the posthuman lens of agential realism. This approach offered us a valuable way to identify the centers and margins of the applications we zoomed in on by analyzing user experience breakdowns and reflecting on the normative practices in the technical development of ML modeling. It also offered a theoretical scaffold to speculate alternatives by enacting different agential cuts and opening up alternative pathways for re-designing the applications. 
Through the case studies, we proposed different ways of cutting the world that provide alternative ways of thinking about not only data and algorithms, but also how to care for ourselves and others. Putting this design implication into action, the new centers we identified in our case studies could be modeled with attention to the sociotechnical milieu of modeling: which materials, discourses, institutions, and power relations are implicated in these new centers? And how should they be related to the old center? 
Future research can explore other posthuman theoretical lenses to identify other ways of decentering. 

Researchers can combine our approach to decentering epistemologies with other emerging practices. For example, an emerging area is \textit{algorithmic auditing} \cite{birhane2024ai}, which identifies and mitigates biases and normative assumptions in AI systems post-development. Researchers advocated that \textit{"auditors should capture and pay attention to what falls outside the measurements and metrics, and to render explicit the assumptions and values the metrics apprehend"} \cite{raji2020closing}. Our proposed approach can complement existing quantitative auditing methods towards more situated qualitative approaches in auditing \cite{raji2020closing, birhane2024ai}. Our analysis echoes the work of \citet{howell_auditing}, which applies a reflective design approach to AI auditing by using each personal experience as ground truth to identify breakdowns in the AI classification system. Our study also expands on this tactic by applying an explicit posthuman orientation and agential realism, which, as we demonstrated, can illuminate alternative design pathways for re-designing the systems being evaluated.

\subsubsection{From Designing with Finished Models to Designing Modeling}

How HCI researchers and interaction designers can effectively work with ML models as a design material is an ongoing open question. Due to the technical complexity and opacity of ML, designers are often left in the position of designing with finished models. With relatively few exceptions (e.g.\cite{sivertsen2024machine,benjamin2021machine}), most HCI literature on this topic treats the complex ML process as a blackbox, and therefore remains at the level of the input and output of the entire system \cite{dove2017ux,yang2020re}. 

Our case studies show the limitations of designing only with finished models. When assumptions about humans are encoded in the worldviews represented through problem formulations in ML modeling, it is very difficult to challenge these or work with alternative ways of representing humans at the level of the interface. These assumptions define which well-being metrics are selected, how data is collected, and which technical process is used to process the data into a model. Indeed, we found that these assumptions in the ML models are mainly inherited and reinforced in the interaction design in our case studies, even when apps are explicitly designed with inclusivity values, such as case studies 1 and 3 \cite{OuraWebsite, ClueMainWS}. For example, case study 1 showed that predictive cycle models require the centering of normative bodies with regular cycles to operate. When using the Clue app, marginalized bodies and bodily experiences may be excluded when it comes to how the app's model aggregates and processes user data. It indicates that when participating in the design of ML models, HCI designers need to be conscious of whether they are taking on the implicit epistemic assumptions and reductive conceptualizations in ML technical practice. 

Our analysis points to the importance for designers to engage in the technical design of ML modeling and use critical and reflective design methods \cite{sengers2005reflective} to {\bf identify the normative biases and assumptions in the models}. While most designers are not trained in the technical aspects of ML, it also represents a unique strength --- as an outsider, designers are particularly well-positioned to spot the implicit assumptions and worldviews taken for granted in the technical fields. Depending on their technical skills, designers can consciously reflect on the values, narratives, and practices in the ML models they use. 

One place for designers to start with is to pay attention to whose {\bf values} are encoded in ML modeling and how. For instance, what performance criteria are used to train the ML model? What type of user well-being behavior is the application fostering? For example, case study 3 brought forward the challenges of ML models to consider the idiosyncrasies of bodily rhythms and 
felt experience of being in transition.
Whilst Oura's readiness model is designed to support recovery from internal and external strains, the assumption of a rhythmic uniformity in bodily transitions induces the hypothesis of normative bodies and daily routines.
Designers can use design methods such as reflective and critical design \cite{sengers2005reflective, odom2024illustrating} to raise awareness of diverse and plural views of bodies and bodily experience in the design qualities of their ML products, and the development team.

For designers with more technical background and AI literacy, they can critically engage with the technical {\bf narratives}. From how ML developers construct narratives of how their ML models work, especially in vernacular terms, researchers can gain insight into what world views and assumptions of AI are embedded \cite{agre2014toward,zhu2009intentional}. Designers can critically reflect on these high-level narratives without necessarily engaging in the technical details. For instance, being part of the interdisciplinary team in case study 2, a designer may notice the divides between ``modeling''  and ``adaptation,''  and between ``current user behavior''  and ``target user behavior'' in how ML developers present their personalization AI. The designer may then apply critical and reflective design methods to consider what is being centered in these dichotomies? What is left to the margins? How do these dichotomies cut the world and the phenomena of human bodies and lived experiences? 
Finally, the growing number of designers with high AI literacy can fully undertake critical technical {\bf practice} in collaboration with ML developers towards more equitable ML modeling techniques.

\subsubsection{From Representing the World, to Being Entangled with the World}

In addition to influencing ML modeling (in)directly, designers can decenter by designing different interactions and relations between the human bodies and their ML models. 
For example, the applications in our case studies treat their ML models of human bodies as objective and (at least eventually) accurate representations of reality. It is important to acknowledge that all models of data ``corrected, rectified, regimented, and in many instances idealized version of the data we gain from immediate observation'' \cite{suppes1966models}. And ML models are no different. Through statistical inference and other mathematical manipulation, ML models present cleaner representations than the messy and complex raw data \cite{suppes1966models,suppes2007statistical}. However, when we incorporate them into real-world applications without acknowledging what is distorted, their underlying assumptions and biases define the centers and cause breakdowns at the margins. 

We suggest that designers should {\bf leverage posthuman theories to explore the design space of alternative relations between ML models and human bodies other than representational ones.} For instance, following diffraction practices \cite{sanches2022diffraction},
design can be used to call attention to the disparity between ML models and human bodies especially in the margins, instead of hiding it. 
Another decentering direction is designing ML applications for a posthuman view of bodies, which does not essentialize bodies to pre-determined categories decided by ML developers. Homewood and colleagues \cite{homewood2021tracing}, when looking at how bodies have been conceptualized in HCI, identify a relational and more-than-human conceptualization of bodies where “ethics and politics are understood as entangled with the material” \cite{homewood2021tracing}. Braidotti proposes a vision of the self as a ``transversal nomadic assemblage''---a fluid, ever-changing collection of experiences, relationships, and interactions \cite{braidotti2019theoretical}--- advocating for localizing the subjects in their social and cultural contexts, and also helping them traverse these \cite{braidotti2014writing}. These posthuman theoretical groundings can provide alternative starting points to conceptualize human bodies in well-being technologies.

Finally, the notion of ``entanglement''  and the impossibility to clearly demarcate or “cleave out” any given entity is also central to posthuman-centered ML modeling, as explored in detail by \citet{frauenberger2019entanglement}. When designing ML well-being technologies entangled in the environment, systems and material infrastructures can result in a reframing of human-centric metrics related to well-being and bodily health (such as those discussed in the case studies) and potentially open up new design approaches. It calls for ML models of individuals to be contestable and porous so that they can adapt to different forms of entanglement. 

\subsection{Challenges Moving Forward}
Adopting a posthuman-centered approach when designing ML technologies for well-being can be valuable in centering marginalized humans and nonhumans implicated by an intended design. However, there are challenges that designers should be aware of when adopting such an approach. Especially when aiming to account for ``all'' through adopting a stance of ``openness'' in a given design space, we should be aware of the potential exclusions that come with such a stance. According to Giraud \cite{giraud2019comes}, not only is openness constituted by particular exclusions as well, but it requires constant work and tinkering. We pose the following question, which we consider valuable for guiding future research on a posthuman-centered ML design stance: \textit{What are potential exclusions that might remain invisible when adopting a posthuman ML approach for designing well-being technologies, and how can we ensure that we do constant work and tinkering to address such exclusions}? In order to do the constant work and tinkering for surfacing what comes after the ``recognition of relationality'' \cite{giraud2019comes}, designers need to develop more tools for working with potential tensions that relate to the ethics and politics of posthuman-centered ML.

\section{Conclusion} 
This paper surfaces the assumptions of the ``human'' embedded in ML well-being technologies, revealing the hidden centers and margins in human-centered ML modeling. We analyzed three case studies of well-being ML technologies: A predictive menstruation and fertility tracker, a social fitness app with personalization ML for motivation; and a self-tracking ring that classifies well-being states. Inspired by critical technical practice, we investigated how the ML features in these apps were experienced to surface breakdowns and identified how narrowly construed assumptions about the ``human'' (body) are encoded as centers in the ML modeling, as well as in the design of these apps. By decentering these identified assumptions through agential realism as an epistemological stance, we proposed alternative centers such as cycle irregularity, human-system entanglements, and uncertain transitions. 

We further speculated towards a posthuman-centered ML modeling approach in well-being technologies with two concrete design implications. First, posthuman-centered ML modeling requires conceptualizing the entangled relationships of
humans and technologies both during the design process and then continuing through the use of the system. Second, posthuman-centered ML modeling requires ML models to directly engage with decentering approaches. Through these contributions, we aim to pave the way towards a much-needed meaningful collaboration between design and machine learning communities. We believe a combination of critical analysis, prospective design and system building, holds the key towards designing deferential and pluralistic technologies for well-being, attentive to the fluidity of bodily expressions, health goals, and ways of being in the world more broadly.

\begin{acks}
This work was partially supported by the Wallenberg AI, Autonomous Systems and Software Program – Humanities and Society (WASP-HS) funded by the Marianne and Marcus Wallenberg Foundation. This work was also partially supported by the Danish Novo Nordisk Foundation under Grant Number NNF20OC0066119. We would like to thank our colleagues and reviewers who provided invaluable feedback. 
\end{acks}

\bibliographystyle{ACM-Reference-Format}
\bibliography{bibliography}

\end{document}